\documentclass[letter-paper]{aa}

\usepackage{graphicx}

\newcommand{\HII}{\ion{H}{II}}

\begin{document}

\title{Massive Star Formation in the W49 Giant Molecular Cloud: \\Implications for the Formation of Massive Star Clusters}

\author{N. L. Homeier \inst{1} \& J. Alves \inst{2}}

\institute{Johns Hopkins University, Department of Physics and Astronomy, USA \and European Southern Observatory, Karl-Schwarzschild Str. 2, 85748 Garching b.  M\"unchen, Germany}

\date{Received / Accepted }

\titlerunning{Massive SF in W49A}
\authorrunning{Homeier \& Alves}

\abstract{We present results from $JHKs$ imaging of the densest
region of the W49 molecular cloud. In a recent paper 
(\cite{AH03}), we reported the detection of (previously unknown)
massive stellar clusters
in the well-known giant radio HII region W49A, and here we 
continue our analysis. 
We use the extensive line-of-sight extinction to isolate a population
of objects associated with W49A. We constrain the slope of the 
stellar luminosity function by constructing an extinction-limited
luminosity function, and use this to obtain a mass function. We find no
evidence for a top-heavy MF, and the slope of the 
derived mass function is $-1.6 \pm 0.3$. We identify candidate 
massive stars from
our color-magnitude diagram, and we use these to estimate the
current total stellar mass of $5-7\times10^{4}$~M$_{\odot}$ in the region
of the W49 molecular cloud covered by our survey. 
Candidate ionizing stars for several
ultra-compact HII regions are detected, with many having multipe candidate 
sources. 
On the global molecular cloud scale in W49, massive star formation apparently 
did not proceed in a single
concentrated burst, but in small groups, or subclusters. This may be 
an essential physical description for star formation in what
will later be termed a 'massive star cluster'.

}
 
\maketitle

\keywords{\HII\ regions --- ISM: individual (W49A) --- open clusters
  and associations: individual (W49A) --- stars: formation --- Galaxy: disk 
--- Stars: winds, outflows}

\section{Introduction}
The W49 Giant Molecular Cloud (GMC) is the most massive in the Galaxy
outside the Galactic center; it extends over more than 50~pc
in diameter (\cite{Setal01}) with a mass of $\sim 10^6$
M$_\odot$. Embedded within this cloud, W49A (\cite{mezger67,SG70}) 
is one of the brightest Galactic giant radio H~II regions 
($\sim 10^7 L_\odot$, \cite{Setal78}). As such, it has been 
used as a template for comparison with extragalactic 
 ``ultradense HII regions'' (UD~\HII; \cite{JK03}), which
appear to be massive star clusters in the process of assembly (\cite{Jetal01,Vetal02}).

The W49A star-forming
region lies in the Galactic plane ($l = 43.17^\circ, b =
+0.00^\circ$) at a distance of 11.4$\pm$1.2 kpc (\cite{Getal92}) and has
$\sim$40 UC~\HII\ regions (e.g.,
\cite{Detal97,depree00,Setal00}) associated with a minimum of that number
of central stars earlier than B3 (later than this, the star does not
put out the necessary UV photons to ionize the surrounding gas, and it
will not be detected as an UC~\HII\ region).  About 12 of these radio
sources are arranged in the well known Welch ``ring'' (\cite{welch87}).
A few other young Galactic clusters have a large number of massive
stars, e.g., the Carina nebula (e.g. \cite{W95,Retal02}), the well-studied
NGC~3603
(e.g. \cite{Metal94,Detal95,Eetal98,Brandletal99,Brandneretal00,Metal02,NPG02,SB04,Stolteetal04}), 
Cygnus OB2 (e.g. \cite{K00}; \cite{Cetal02}; 
\cite{H03}), the Arches cluster 
(e.g. \cite{SSF98,Blumetal01b,Figeretal02,Setal02}), and
Westerlund~1 (\cite{CN02}), or
are very young, e.g. NGC~3567 (\cite{Betal03,Fetal02}), W42 
(\cite{Betal00}),
and W31 (\cite{KK02,Betal01a}) but no other known
region has a large number of massive stars in such a highly embedded
and early evolutionary state.  For this reason W49A is unique in our
known Galaxy. 
 
To uncover the embedded stellar population in W49A we
performed a $5^\prime\times5^\prime$ (16 pc $\times$ 16 pc),
deep J, H, and Ks-band imaging survey centered on the densest region
of the W49 GMC (\cite{Setal01}, see their Figure 2). The initial
results were presented in Alves \& Homeier (2003), where we reported
the discovery of one massive and three smaller stellar clusters
detected at NIR wavelengths. In this companion paper
we present our photometric results, including the number of massive star
candidates, objects with infrared excesses, and candidate ionizing
sources of compact and ultracompact \HII\ regions. 

%Out of the giant radio \HII\ regions so far studied, W49A was thought
%to be the only one without an older, relatively unembedded massive 
%stellar population. 
%A series of investigations of giant radio \HII\ regions 
%by Blum et al. (1999, 2000, 2001)

%The paper is organized as follows. In \S2 we briefly review the stages
%of massive star formation, in \S3 we discuss our observations and 
%reductions, in \S4 we present our photometric results, in \S5 we discuss 
%these results in the context of previous observations, and in \S6 we present
%our conclusions.

\section{Stages of massive star formation}

To better interpret what we observe in the W49A star-forming region, 
we will briefly mention a simplified version of 
the stages of massive star formation. 
The hot core phase is that of a rapidly 
accreting, massive protostar. Although the protostar is emitting 
UV photons at this stage, the H~II emission is ``quenched''
due to the high accretion rate (Walmsley~1995, Churchwell~2002).
The next phase is the ultra-compact H~II (UC~\HII) phase, and is the
 most observationally well-studied. An UC~\HII\ region contains
 a central hydrogen-burning star which has ceased appreciable 
 accretion.  The star's UV flux eats through its gas and dust cocoon and 
will eventually break out of the dense local medium to ionize
surrounding more diffuse ISM. UC~\HII\ regions are radio-, far-IR-, and 
sometimes mid-IR-bright, but often undetectable at NIR
wavelengths due to high local extinction. As the star disperses
 more of the surrounding gas and dust, the UC~\HII\ region
 becomes observable at shorter and shorter wavelengths, until the
 central object finally emerges as an unobscured massive star (see 
Churchwell~2002).

%%%%%%%%%%%%%%%%%%%%%%%%%%%%%%%%%%%%%%%%%%%%%%%%%%%%%%%%%%%%%%%%%%%%%%%%%
%%%%%%%%%%%%%%%%%%%%%%%%%%%%%%%%%%%%%%%%%%%%%%%%%%%%%%%%%%%%%%%%%%%%%%%%%
\begin{figure*}[t]
\centering
\includegraphics[scale=.45]{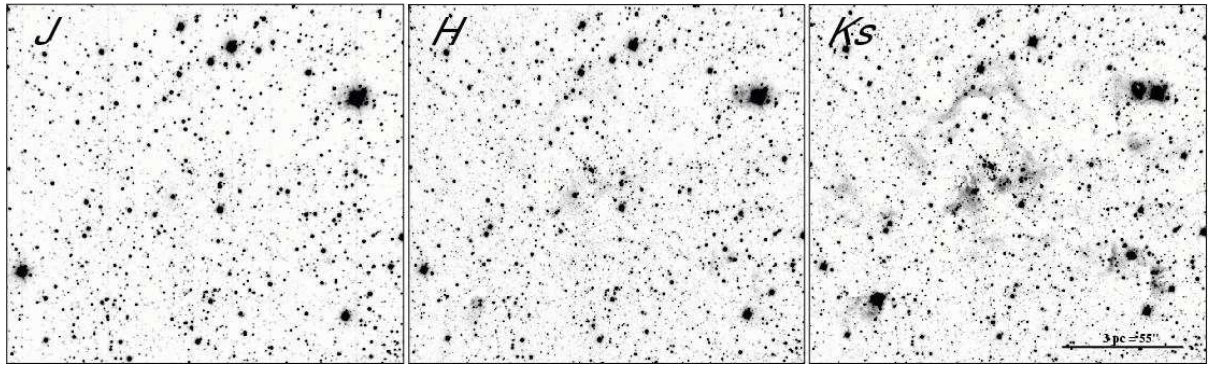}
\caption{$J, H,$ and $Ks$ images of the W49A region. The main star cluster
begins to appear in the $H$ image, but is readily apparent in the $Ks$ image,
as are many diffuse nebular features.
North is up and East to the left.
\label{triple}}
\end{figure*}
%%%%%%%%%%%%%%%%%%%%%%%%%%%%%%%%%%%%%%%%%%%%%%%%%%%%%%%%%%%%%%%%%%%%%%%%%
%%%%%%%%%%%%%%%%%%%%%%%%%%%%%%%%%%%%%%%%%%%%%%%%%%%%%%%%%%%%%%%%%%%%%%%%%

\section{Observations and Data Reduction}\label{sec:observations}
The observations were taken June 8, 2001, with the SOFI 
(\cite{Moorwoodetal98}) near-infrared
 camera on the ESO's 3.5m New Technology Telescope (NTT) on La Silla,
 Chile, during a spell of good weather and exceptional seeing (FWHM
 of the final combined images
 $\sim$ 0.5-0.7$^{\prime\prime}$). SOFI employs a $1024\times1024$
 Hawaii HgCdTe array, and the observations were taken with a plate scale
 of $0.288\arcsec/$pixel. A set of 30 dithered images of 60
 seconds each were taken in the J, H, and Ks filters. The images were
 combined with the DIMSUM\footnote{DIMSUM is the Deep Infrared
 Mosaicing Software package developed by Peter Eisenhardt, Mark
 Dickinson, Adam Stanford, and John Ward, and is available via ftp
 from ftp://iraf.noao.edu/contrib/dimsumV3/.} package. The pixel scale
of the final combined images is $0.15\arcsec/$pixel. Standard stars 
 9136, 9157, and 9172 from the Persson catalog (Persson et~al. 1998)
 were observed at the beginning, middle, and end of the night
 to obtain an airmass solution. Photometry was
 performed with the DAOPHOT package in IRAF\footnote{IRAF is
 distributed by the National Optical Astronomy Observatories}.
 DAOFIND was used to detect sources above a threshold of 5 sigma. 
 PSF models were constructed for each image using $5-7$ isolated, bright
 stars and the tasks PSTSELECT and PSF, and ALLSTAR were run to extract
 the final photometry and aperture corrections were performed. 
 $J$ and $H$ coordinates were transformed to $Ks$ image
 coordinates using GEOMAP and GEOXYTRAN.
 Objects with errors larger than 0.15 magnitudes and objects within 150
 pixels of the image edges were excluded. Our final $J$, $H$, and $Ks$ images 
 are shown in Figure\ref{triple}. Our final samples contain 2255 and
 7299, for the matched $JHKs$ and $HKs$ lists, respectively.

\subsection{Comparison with Previous Observations}

We compared our $Ks$ photometry with the $K$ observations of 
Conti~\&~Blum~(2002).
From a sample of 493 stars in both data sets, we find an offset of 
0.1 magnitudes between the $K$ and $Ks$ magnitudes, with the SOFI photometry
presented here being 0.1 magnitudes fainter than the OSIRIS photometry. 
This offset can be accounted for by the different filter response
curves and the highly reddened nature of the objects. 

\begin{figure}
\includegraphics[scale=.5]{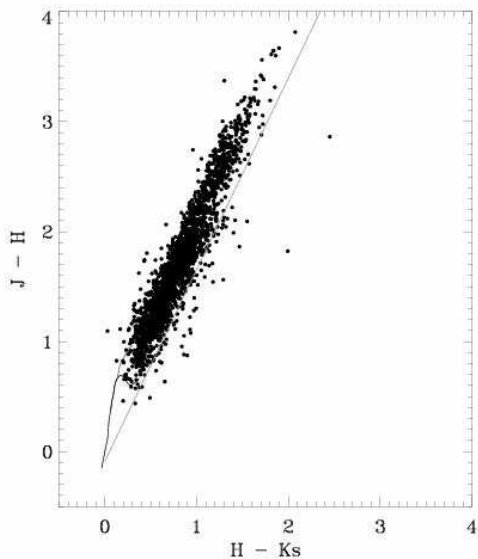}
\caption{Color-color diagram for our survey. The main sequence and giant
tracks are overplotted, as is a light solid line indicating the reddening
sequence for the bluest, hottest stars. A Rieke~\&~Lebofsky (1985) 
extinction law was used.
  \label{fig:cc}}
\end{figure}

We performed tests
with the SYNPHOT task CALCPHOT for the SOFI $Ks$ and OSIRIS $K$ 
($K_{185}$) filters. Using the Galactic extinction law of Clayton, Cardelli, \&
Mathis (CCM) (1989) for $E(B-V)=10.85$ ($A_{V}\sim30$), 
SOFI $H-Ks=1.561$ for a 30000K blackbody,
there is a $0.037-0.056$ magnitude difference for $30000-3000$K blackbodies,
with the SOFI photometry being fainter. With $E(B-V)=7.0$ ($A_{V}\sim20$), 
SOFI $H-Ks=0.832$ for a 30000K blackbody, 
there is a $0.014-0.029$ magnitude difference for $30000-3000$K blackbodies,
again with the SOFI photometry being fainter. It would seem that
approximately $0.04-0.09$ magnitudes are unaccounted for, however, the extinction
laws available with SYNPHOT do not include the widely accepted 
Rieke \& Lebofsky (RL) (1985) Galactic extinction law, which we use 
throughout the rest of the paper. 
The slope of this extinction law also describes the slope
of our color-color relation shown in Figure~\ref{fig:cc}.

For $A_{K}=1$, the CCM law gives $H-K$=0.6, while the RL law
gives $H-K=0.57$. Thus there is a difference of 
$0.03-0.09$ magnitudes for $A_{K}=1-3$, typical of the stars in this 
region. Therefore, if the RL extinction law (and not the CCM law) 
accurately describes the extinction
along the line of sight to W49A (which we have evidence for) 
then the 0.1 magnitude offset between our SOFI $Ks$
magnitudes and the $K$ magnitudes of Conti \& Blum (2002) 
should be due to the different filter response curves and the
highly reddened nature of the stars.

%%%%%%%%%%%%%%%%%%%%%%%%%%%%%%%%%%%%%%%%%%%%%%%%%%%%%%%%%%%%%%%%%%%%%%%%%
%%%%%%%%%%%%%%%%%%%%%%%%%%%%%%%%%%%%%%%%%%%%%%%%%%%%%%%%%%%%%%%%%%%%%%%%%
%%%%%%%%%%%%%%%%%%%%%%%%%%%%%%%%%%%%%%%%%%%%%%%%%%%%%%%%%%%%%%%%%%%%%%%%%
%%%%%%%%%%%%%%%%%%%%%%%%%%%%%%%%%%%%%%%%%%%%%%%%%%%%%%%%%%%%%%%%%%%%%%%%%

\subsection{Completeness limits}\label{sec:completeness-limit}

 The completeness limits were determined by adding 500-1500 fake stars to
 each image and extracting them in the same way as the data analysis
 was performed. The fake stars were created using the psf image used
 with the ALLSTAR task, and input to make images using ADDSTAR. 
We consider a star as recovered only 
if its recovered magnitude is within 
0.15 magnitudes (our error cut) of the input magnitude. 
The 80 \% completeness limits for the  $J$, $H$, and $Ks$ filter images are 
20.0, 
18.7, and 17.2, respectively. The limits reflect the increasing importance
of crowding in our images from $J$ to $Ks$.

Due to crowding concerns, we also performed completeness 
tests on the central $500\times500$ pixels of our images. 
For the $H$ and $Ks$ images, the 80\% completeness limits were approximately
0.5 magnitudes brighter than the limits for the entire field. The $J$ limit
was unaffected.

\section{Results}

\subsection{Luminosity Functions}
\label{sec:LF}

Our images contain many stars along the line of sight, but we can
use the reddening within the Galactic disk to our advantage.
We attempt to identify a stellar population associated with the W49A 
region by first selecting objects with $H-Ks$
colors red enough to be consistent with a distance of $11.4$~kpc along the
Galactic plane. This can be calculated by assuming an
exponential distribution of dust (as in \cite{Hetal03}) 
so that the extinction follows the form:

a$_{K}$(R)=a$_{K,0}$e$^{-(R-R_{0})/\alpha_R}$, and 

$R=(x^{2}+R^{2} _{0}-2xRcosl)^{1/2}$, 

\noindent where $x$ is $11.4\pm1.2$~kpc from Gwinn~et~al.~(1992), $l=43$, 
R$_{0}=8$~kpc, and $\alpha_{R}$=3.0~kpc (\cite{Ketal91}).
.
For these parameters we arrive at A$_{K}=2.1$ and $H-K=1.2$ for a 
Rieke~\&~Lebofsky~(1985) reddening law.

In the remainder of the paper, we consider objects within $45\arcsec$ of 
19:10:17.5, 09:06:21 (J2000)
to be associated with Cluster 1, which corresponds to the arc of 
ionized emission to the North, and a physical distance of 2.5~pc. 

%%%%%%%%%%%%%%%%%%%%%%%%%%%%%%%%%%%%%%%%%%%%%%%%%%%%%%%%%%%%%%%%%%%%%%%%%
%%%%%%%%%%%%%%%%%%%%%%%%%%%%%%%%%%%%%%%%%%%%%%%%%%%%%%%%%%%%%%%%%%%%%%%%%
\begin{figure}
\includegraphics[scale=.5]{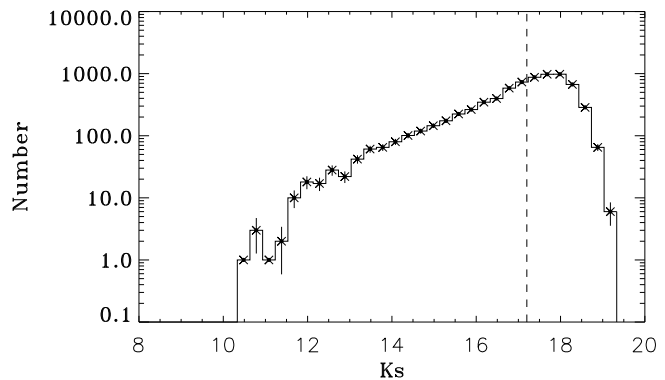}
\caption{The K-band luminosity function for all objects. The binsize is 0.3 
magnitudes. We show the 80\% completeness limit as a dashed line.}
\label{fig:lfall}
\end{figure}
%%%%%%%%%%%%%%%%%%%%%%%%%%%%%%%%%%%%%%%%%%%%%%%%%%%%%%%%%%%%%%%%%%%%%%%%%
%%%%%%%%%%%%%%%%%%%%%%%%%%%%%%%%%%%%%%%%%%%%%%%%%%%%%%%%%%%%%%%%%%%%%%%%%

%%%%%%%%%%%%%%%%%%%%%%%%%%%%%%%%%%%%%%%%%%%%%%%%%%%%%%%%%%%%%%%%%%%%%%%%%
%%%%%%%%%%%%%%%%%%%%%%%%%%%%%%%%%%%%%%%%%%%%%%%%%%%%%%%%%%%%%%%%%%%%%%%%%
\begin{figure}
\includegraphics[scale=0.5]{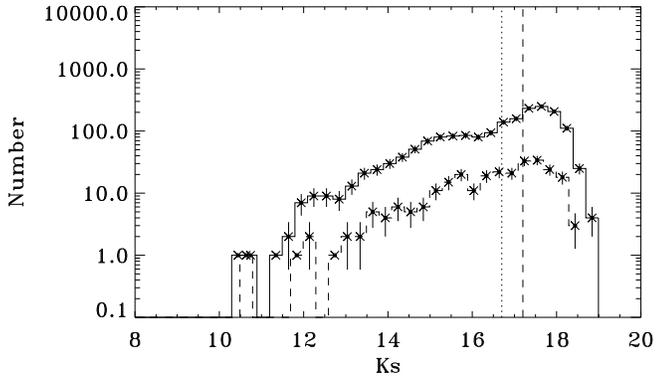}
\caption{The K-band luminosity function for objects with $H-K \ge 1.2$ which
are likely to be associated with the W49A star-forming region
(see text).
The binsize is 0.3 magnitudes. The dashed histogram indicates objects within 
$45\arcsec$ of our adopted center of Cluster~1. We include the 80\% 
completeness
limit for the entire field as a dashed line, and for the inner $500\times500$
pixels as a dotted line.}
\label{fig:lfred}
\end{figure}
%%%%%%%%%%%%%%%%%%%%%%%%%%%%%%%%%%%%%%%%%%%%%%%%%%%%%%%%%%%%%%%%%%%%%%%%%
%%%%%%%%%%%%%%%%%%%%%%%%%%%%%%%%%%%%%%%%%%%%%%%%%%%%%%%%%%%%%%%%%%%%%%%%%

%%%%%%%%%%%%%%%%%%%%%%%%%%%%%%%%%%%%%%%%%%%%%%%%%%%%%%%%%%%%%%%%%%%%%%%%%
%%%%%%%%%%%%%%%%%%%%%%%%%%%%%%%%%%%%%%%%%%%%%%%%%%%%%%%%%%%%%%%%%%%%%%%%%
\begin{figure}
\includegraphics[scale=.5]{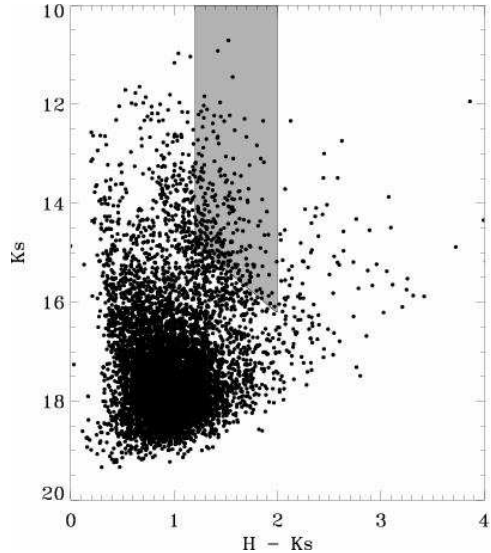}
\caption{$(H-Ks)$ Color-Magnitude diagram showing the limits for an
extinction-limited sample. We chose an $H-K=2.0$ limit to cover a 
reasonable range of reddening. For the entire field, our 80 \%
completeness limits are $H=18.7$ and $Ks=17.2$. However, crowding in the
center of the image reduces this to $H=18.2, Ks=16.7$  Thus at $Ks=16.2$, 
$H-K=2$, we are above 80 \% completeness for everything brighter and 
bluer than these limits. This defines the bottom-right corner of 
the overplotted polygon. The slope of the bottom edge is determing by the 
extinction law, and the $H-K=1.2$ limit is described in \S~\ref{sec:LF}.
  \label{fig:cmd_extinclim}}
\end{figure}
%%%%%%%%%%%%%%%%%%%%%%%%%%%%%%%%%%%%%%%%%%%%%%%%%%%%%%%%%%%%%%%%%%%%%%%%%
%%%%%%%%%%%%%%%%%%%%%%%%%%%%%%%%%%%%%%%%%%%%%%%%%%%%%%%%%%%%%%%%%%%%%%%%%

%%%%%%%%%%%%%%%%%%%%%%%%%%%%%%%%%%%%%%%%%%%%%%%%%%%%%%%%%%%%%%%%%%%%%%%%%
%%%%%%%%%%%%%%%%%%%%%%%%%%%%%%%%%%%%%%%%%%%%%%%%%%%%%%%%%%%%%%%%%%%%%%%%%
\begin{figure}
\includegraphics[scale=.5]{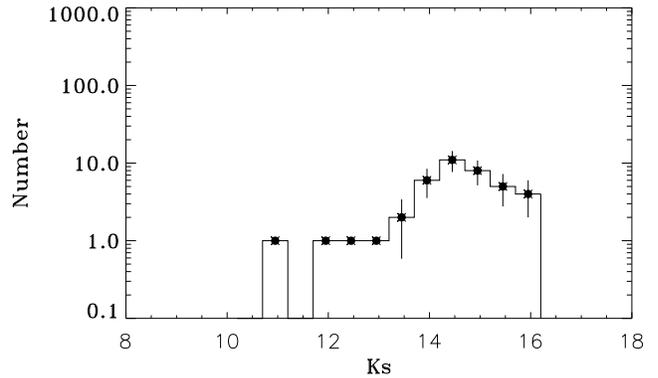}
\caption{Extinction-limited K-band luminosity function for Cluster 1. The binsize 
is 0.5 magnitudes. }
\label{fig:extinclimclus1}
\end{figure}
%%%%%%%%%%%%%%%%%%%%%%%%%%%%%%%%%%%%%%%%%%%%%%%%%%%%%%%%%%%%%%%%%%%%%%%%%
%%%%%%%%%%%%%%%%%%%%%%%%%%%%%%%%%%%%%%%%%%%%%%%%%%%%%%%%%%%%%%%%%%%%%%%%%

%%%%%%%%%%%%%%%%%%%%%%%%%%%%%%%%%%%%%%%%%%%%%%%%%%%%%%%%%%%%%%%%%%%%%%%%%
%%%%%%%%%%%%%%%%%%%%%%%%%%%%%%%%%%%%%%%%%%%%%%%%%%%%%%%%%%%%%%%%%%%%%%%%%
\begin{figure}
\includegraphics[scale=.5]{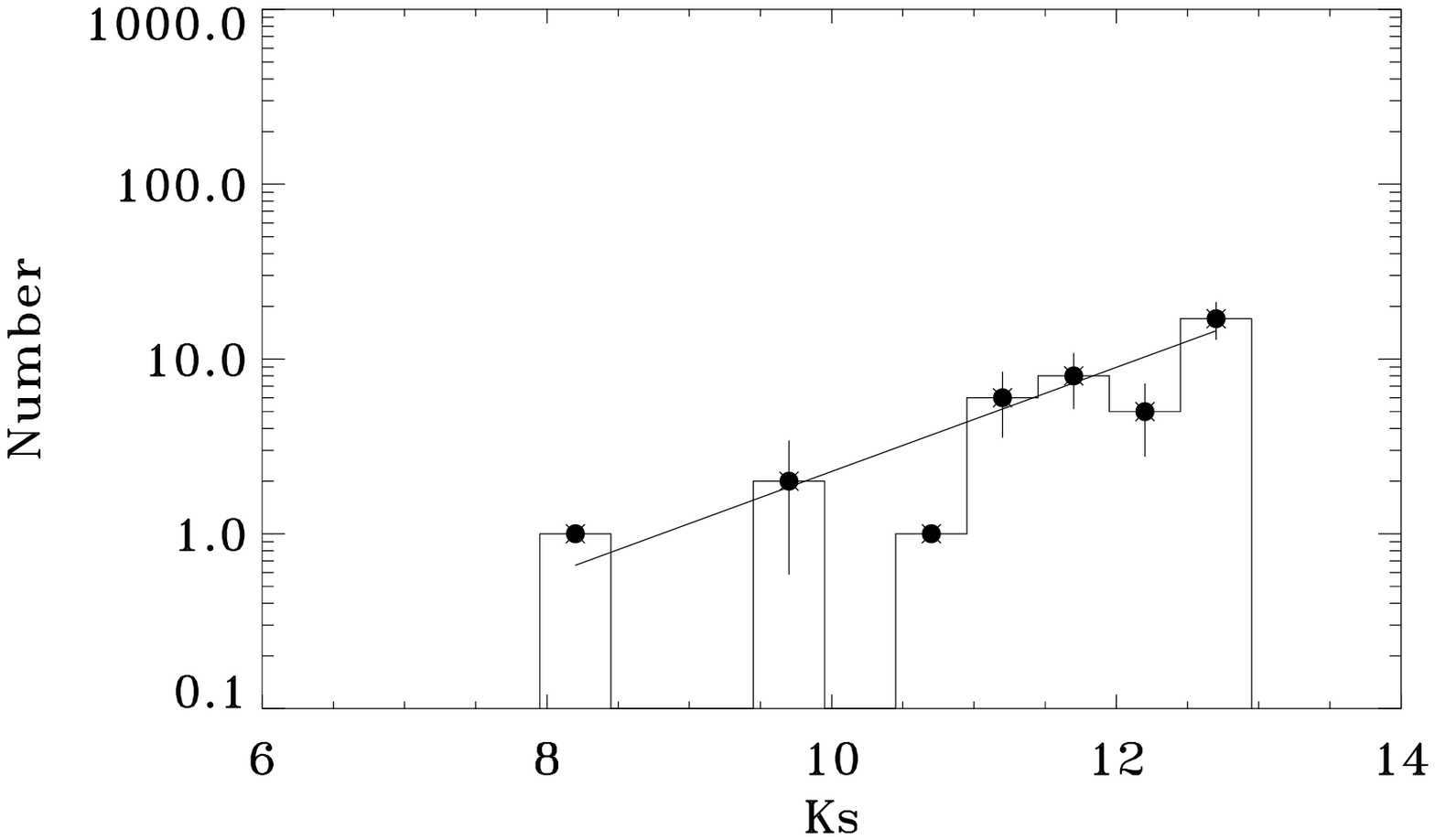}
\caption{Extinction-limited extinction-corrected K-band luminosity function 
for Cluster 1. The binsize is 0.5 magnitudes. The slope of the fitted line 
is $0.30 \pm 0.16$. We used a larger bin size due to errors in extinction
correction caused by intrinsic $H-Ks$ colors.}
\label{fig:extinclimclus1dered}
\end{figure}
%%%%%%%%%%%%%%%%%%%%%%%%%%%%%%%%%%%%%%%%%%%%%%%%%%%%%%%%%%%%%%%%%%%%%%%%%
%%%%%%%%%%%%%%%%%%%%%%%%%%%%%%%%%%%%%%%%%%%%%%%%%%%%%%%%%%%%%%%%%%%%%%%%%

%%%%%%%%%%%%%%%%%%%%%%%%%%%%%%%%%%%%%%%%%%%%%%%%%%%%%%%%%%%%%%%%%%%%%%%%%
%%%%%%%%%%%%%%%%%%%%%%%%%%%%%%%%%%%%%%%%%%%%%%%%%%%%%%%%%%%%%%%%%%%%%%%%%
\begin{figure}
\includegraphics[scale=.5]{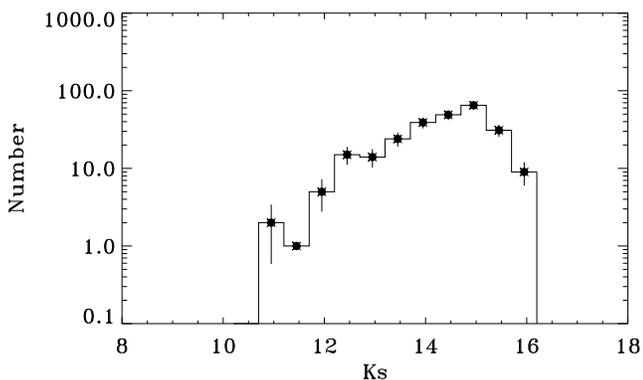}
\caption{Extinction-limited K-band luminosity function for all objects. 
The binsize is 0.5 magnitudes.  
\label{fig:extinclimall}}
\end{figure}
%%%%%%%%%%%%%%%%%%%%%%%%%%%%%%%%%%%%%%%%%%%%%%%%%%%%%%%%%%%%%%%%%%%%%%%%%
%%%%%%%%%%%%%%%%%%%%%%%%%%%%%%%%%%%%%%%%%%%%%%%%%%%%%%%%%%%%%%%%%%%%%%%%%

%%%%%%%%%%%%%%%%%%%%%%%%%%%%%%%%%%%%%%%%%%%%%%%%%%%%%%%%%%%%%%%%%%%%%%%%%
%%%%%%%%%%%%%%%%%%%%%%%%%%%%%%%%%%%%%%%%%%%%%%%%%%%%%%%%%%%%%%%%%%%%%%%%%
\begin{figure}
\includegraphics[scale=.5]{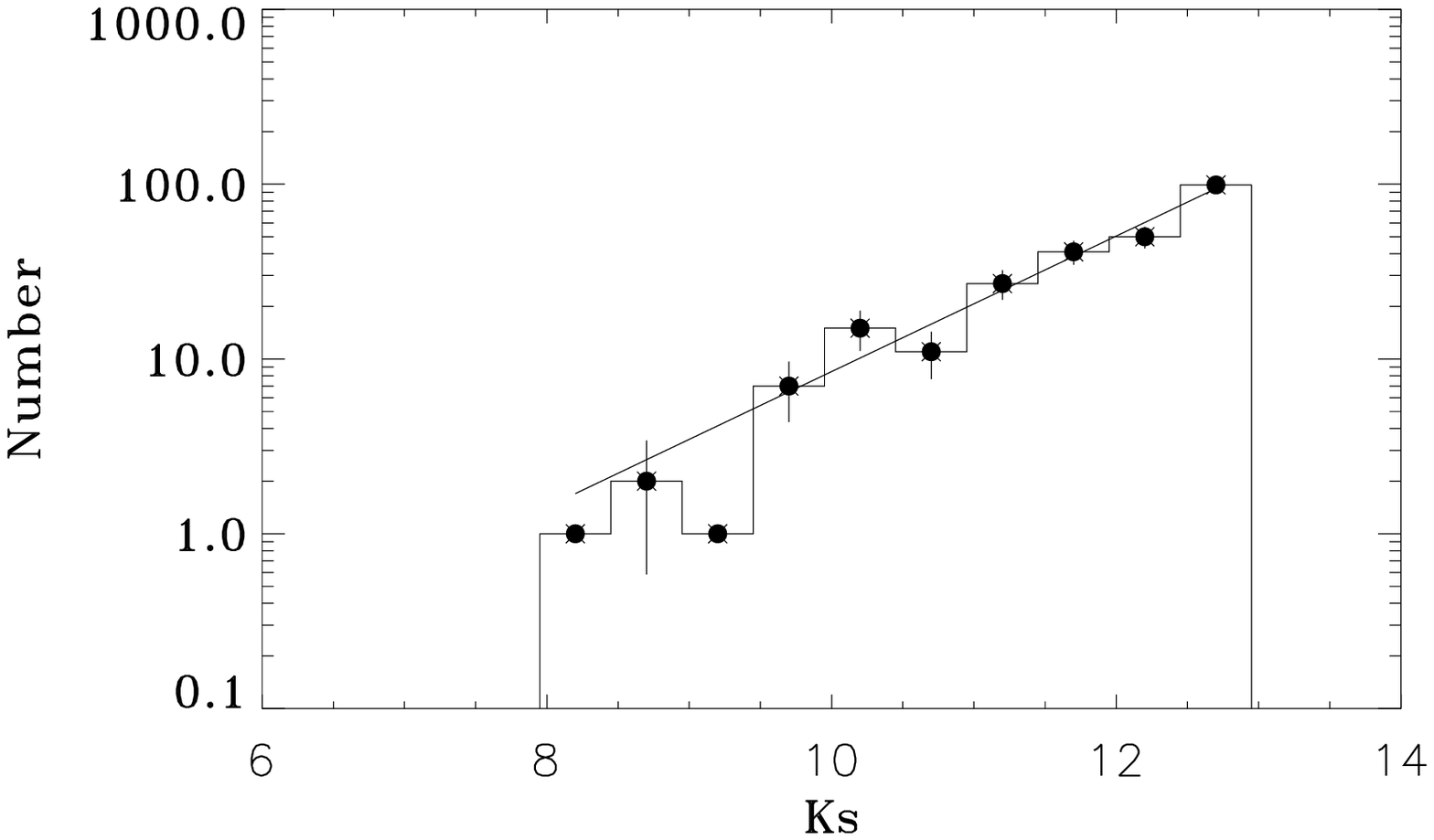}
\caption{Extinction-limited extinction-corrected K-band luminosity function 
for all objects. The binsize is 0.5 magnitudes. The slope of the fitted line 
is $0.37 \pm 0.07$. We used a larger bin size due to errors in extinction
correction caused by intrinsic $H-Ks$ colors.}
\label{fig:extinclimalldered}
\end{figure}
%%%%%%%%%%%%%%%%%%%%%%%%%%%%%%%%%%%%%%%%%%%%%%%%%%%%%%%%%%%%%%%%%%%%%%%%%
%%%%%%%%%%%%%%%%%%%%%%%%%%%%%%%%%%%%%%%%%%%%%%%%%%%%%%%%%%%%%%%%%%%%%%%%%

In Figure \ref{fig:lfall} we present the K-band luminosity function for all
objects in our $H$ and K sample, and in Figure \ref{fig:lfred} we select only 
those objects with $H-K \ge 1.2$ as being located at or farther than the W49 
molecular cloud as described above. 
The histogram for objects within $45 \arcsec$ of Cluster 1 is plotted with a
dashed line, and a solid line indicates all objects outside of this region.
The binsize is 0.3 magnitudes, and the bin boundary at the faint magnitude
limit was chosen to be the magnitude of the faintest star in each sample.

We would like an unbiased luminosity function for the stars associated with 
W49, so we select an extinction-limited sample of stars 
within $45\arcsec$ of our adopted center of Cluster~1. We expect negligible 
background contamination near Cluster~1 due to the large optical depth of
the W49 modelcular cloud. The magnitude and 
color limits of our 
extinction-limited sample are shown in Figure \ref{fig:cmd_extinclim}. The
$H-Ks=1.2$ limit is set by our best estimate of foreground extinction as
previously described. 
Our 80\% completeness limits for the entire field are at 
$H=18.7$ and $Ks=17.2$. Thus at $Ks=16.7$, $H-Ks=2$, we are
above 80\% completeness for everything brighter and bluer than these 
limits. However, crowding in the center reduces these 80\% 
completeness limits to $H=18.2$ and $Ks=16.7$, and thus $Ks=16.2, H-Ks=2$
defines the bottom-right 
corner of the overplotted region. The slope of the bottom edge is 
determined by a Rieke \& Lebofsky extinction law (1985).

The extinction-limited KLF for Cluster~1 is shown in 
Figure~\ref{fig:extinclimclus1},
and for everything in our field 
Figure~\ref{fig:extinclimall}. For these histograms, the bin boundary on 
the faint end was set at the faintest star in each sample. 
The error bars are $\sqrt{N}$ for the number of stars in each bin.

Since both samples suffer from severe non-uniform extinction, we 
corrected for this effect assuming an intrinsic color of $H-Ks=0.15$. 
This was chosen based on the knowledge 
that all stars without hot dust are intrinsically 
nearly colorless in the near-infrared, with $H-Ks$ ranging from 0.0 to 0.3. 
We expect objects associated with the W49A star-forming region to be early-type
stars with intrinsic $H-Ks$ near 0.0, whereas giant stars should
have intrinsic $H-Ks$ up to 0.3. Thus we calculate extinction-corrected  
K magnitudes as $K_{corr}=K_{obs} + (($H-Ks-0.15$)/0.57)$.
Dereddening a star with an intrinsic color $H-Ks=0$ to $H-Ks=0.15$ will result
in an inferred $Ks$ magnitude which is 0.25 magnitudes too faint, whereas a
star with an intrinsic $H-Ks=0.3$ will be 0.25 magnitudes too bright.

Our extinction-corrected extinction-limited KLFs for Cluster~1 and
for all objects in our field are shown in 
Figures~\ref{fig:extinclimclus1dered} and \ref{fig:extinclimalldered}.
We use 0.5 magnitude bins for these samples, due to the
uncertainty in the extinction correction. The bin boundary at the faint
end was set to K=12.95, the faintest extinction-corrected magnitude allowed
by our selection criteria. 
A linear least-squares fit was performed, yielding a slope of $0.30 \pm 0.16$ 
for Cluster~1 and $0.37 \pm 0.07$ for the entire field. 

\subsection{Mass Function}

\begin{figure*}
\includegraphics[scale=.5]{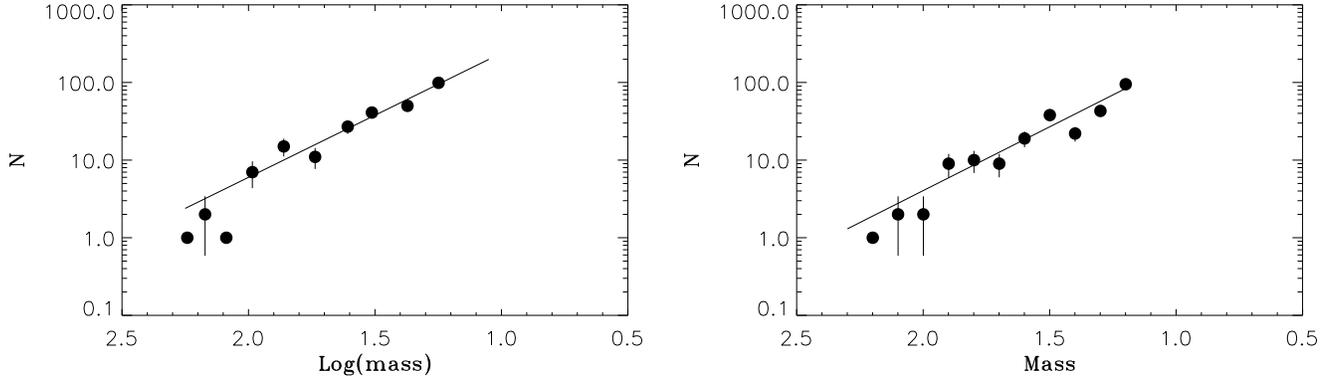}
\caption{Mass functions from the extinction-limited extinction-corrected sample
shown in Figure \ref{fig:extinclimalldered}. In the left panel, we transformed 
the $Ks$ magnitudes
of the luminosity function into mass. The mass bin sizes thus depend on the
mass-magnitude relationship.
The fitted slope is $-1.60 \pm 0.28$. In the right panel, masses were derived 
based 
on the $Ks$ magnitudes for individual stars, and the log(M) bins are 
0.1. The fitted slope is $-1.64\pm 0.28$. 
We do not find
evidence for a top-heavy IMF. 
\label{fig:mf}}
\end{figure*}

Any photometric mass function relies on a 
magnitude-mass relation, which has its source in a luminosity-mass
relation. 
We take the relationship between initial mass and absolute
$K$ magnitude from the $4\times10^{5}$~yr isochrones of 
Lejeune~\&~Schaerer (2001) with enhanced mass loss rates. 
We can then construct a mass function
by converting our extinction-limited extinction-corrected 
$Ks$ luminosity function for our entire field by transforming
each magnitude bin to a mass bin. We can also
convert magnitudes for individual stars into masses,
then bin these masses to arrive at a mass histogram.
The mass functions derived in these two ways are
shown in Figure~\ref{fig:mf}.
We extrapolated the magnitude-mass relation to infer masses for the most
luminous stars, which are more luminous than the 120~M$_{\odot}$ models.
Errors of $10-20$\% in the mass estimate are expected simply
from the uncertainty in the distance.

Our slopes are derived from linear least-squares fits weighted with the
errors bars derived from Poisson statistics ($\sqrt{N}$).
The mass function slopes yielded by the two methods, $-1.60\pm 0.28$ and 
$-1.64\pm 0.28$, are in excellent agreement
The error in each slope measurement is large and there are many sources of 
uncertainty, but we can conclude that we do not find
evidence for a top-heavy IMF. 
If we use the 1~Myr isochrones, our measured slopes are $-1.70\pm 0.30$ 
and $-1.57\pm 0.31$, within the $1\sigma$ uncertainties.

The slope of our mass function and the number of stars in the sample
indicates that we should have at least one
M$> 200$~M$_{\odot}$ in our extinction-limited
sample. Our $J$ image does not go deep enough for us to securely identify
extremely massive candidates. There are several luminous objects at $Ks$
for which we lack $J$ magnitudes, and we are therefore unable to 
quantify the contribution to the $Ks$ magnitude from hot dust.

%%%%%%%%%%%%%%%%%%%%%%%%%%%%%%%%%%%%%%%%%%%%%%%%%%%%%%%%%%%%%%%%%%%%%%%%%
%%%%%%%%%%%%%%%%%%%%%%%%%%%%%%%%%%%%%%%%%%%%%%%%%%%%%%%%%%%%%%%%%%%%%%%%%
%%%%%%%%%%%%%%%%%%%%%%%%%%%%%%%%%%%%%%%%%%%%%%%%%%%%%%%%%%%%%%%%%%%%%%%%%
%%%%%%%%%%%%%%%%%%%%%%%%%%%%%%%%%%%%%%%%%%%%%%%%%%%%%%%%%%%%%%%%%%%%%%%%%

\subsection{Candidate massive stars}

%%%%%%%%%%%%%%%%%%%%%%%%%%%%%%%%%%%%%%%%%%%%%%%%%%%%%%%%%%%%%%%%%%%%%%%%%
%%%%%%%%%%%%%%%%%%%%%%%%%%%%%%%%%%%%%%%%%%%%%%%%%%%%%%%%%%%%%%%%%%%%%%%%%
\begin{figure}
\includegraphics[scale=.5]{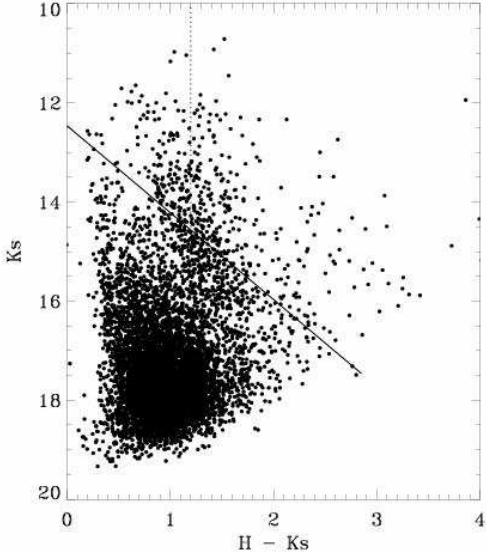}
\caption{$(H-K)$ Color-Magnitude diagram showing the limits for identifying
candidate massive stars. The solid line represents the reddening line for 
a $4\times10^{5}$ yr 20 M$_{\odot}$ star (\cite{LS01}), 
from A$_{K}=0$ to A$_{K}=5$. 
The dotted line represents the $H-K=1.2$ limit described in \S~4.1.
  \label{fig:cmd_massive}}
\end{figure}
%%%%%%%%%%%%%%%%%%%%%%%%%%%%%%%%%%%%%%%%%%%%%%%%%%%%%%%%%%%%%%%%%%%%%%%%%
%%%%%%%%%%%%%%%%%%%%%%%%%%%%%%%%%%%%%%%%%%%%%%%%%%%%%%%%%%%%%%%%%%%%%%%%%

To estimate the number of stars with masses $\ge 20$~M$_{\odot}$ 
associated with the W49A region, 
% 15.28^{+0.22}_{-0.24}
we will assume
intrinsic colors of $H-K=0$ and calculate the unobscured apparent K magnitude 
as K$_{obs}+(0.57/H-K)$. As in the previous section, we use the 
relation between mass and absolute K magnitude from the
Lejeune \& Schaerer (2001) models at $4\times10^{5}$~yr for solar metallicity and
enhanced mass loss. Assuming an age from $3\times 10^{5}$ yr to 2~Myr has a 
negligible effect on our overall result. 

Figure \ref{fig:cmd_massive} shows our CMD with the 
position of a $4\times10^{5}$~yr
20~M$_{\odot}$ star, and a reddening line indicating A$_{K}=5$. All stars 
above this line with $H-K \ge 1.2$ are identified as candidate massive stars.
We will use this sample later to estimate the total stellar mass in the region.

%In Table~\ref{tab:ostars} we present coordinates, magnitudes, and colors
%for these objects.

\subsection{Contamination}

There is no robust
way to measure the background in such a region, as the cluster is 
embedded in a molecular cloud, which means the extinction is non-uniform
across the field. However, one likely contaminant is disk giant stars.
Absolute $Ks$ magnitudes for the brightest disk giant stars should 
be $\sim -4.7$ (\cite{SG00}), 
which is equivalent to an apparent magnitude of $\sim 10.1$  at a distance 
of 12~kpc (neglecting extinction), just farther than the W49A region. 
Assuming $A_{K}=2-2.5$ magnitudes, they would have apparent magnitudes
of $\sim 12.1-12.6$. If we select stars with $H-Ks >1.2$ and $Ks$ in this 
magnitude range, we find that they are not uniformly
distributed over our field, but fall preferentially on the northern half.
This is consistent with their identification as background giant stars, as
the W49 molecular cloud is less dense as one goes from the center
to the northern edge of the image. From the nonuniform distribution 
of reddended sources in our field, which we identify as background giants,
we estimate 
that they contribute $20-30$ stars to our total. 
Another source of contamination in our census of massive stars
is a possible population of stars which are undetected at $J$ with 
$Ks$ excesses, which could make 
some less massive stars appear as more massive stars. 

\subsection{Objects with infrared excesses}

Our $J-H$ vs. $H-Ks$ color--color diagram is shown in Figure \ref{fig:cc},
with the main sequence and giant tracks overplotted as solid and dotted
lines (\cite{BB88}). The reddening boundary for the hottest stars is 
plotted as a
dashed line. We can see that several stars fall to the right of this 
line, which would indicate an $H-Ks$ color which is affected by not only
extinction, but a $Ks$ excess due to hot dust. However, most of these 
are not extincted 
enough to be part of the W49 region and also fall near the edges of 
our images. These could be photometric outliers, or true Ks excess
objects along the line of sight, most probably, the population is a 
combination of the two. Only two stars with strong $Ks$ excesses 
are likely to be part of the W49A region. Both of these are within 2~pc of the 
projected center of Cluster~1. One is faint and appears to have an unresolved
companion at K, suggesting the result could be due to a deblending error. 
The other was identified as
star no. 2 by Conti \& Blum (2002). This object has an excess of approximately
0.7 magnitudes and thus from its corrected color and magnitude appears to be 
a star with a mass 20-25~M$_{\odot}$. It is located between the projected 
center of Cluster 1 and the ring of ultracompact HII\ regions. This star is 
an obvious candidate for follow--up observations looking for evidence
that the hot dust which surrounds the star is in the shape
of a remnant accretion disk.

\subsection{Candidate ionizing sources of compact and ultracompact HII\ 
regions} 

Conti \& Blum (2002) detected two UC\HII\ regions in their $H$ and $K$ images,
radio sources 'F' and 'J2'. We also detect these sources, and 
in Table~\ref{tab:uchii} we list candidate ionizing sources for 
the compact and ultracompact
\HII\ regions CC, F, J2, R and Q, S, and W49 South (names from \cite{Detal97}). 
For regions with multiple detections, we have selected only objects with 
inferred M $> 15$~M$_{\odot}$. We list inferred masses in column 9, and
we note that with the relation we are using, five objects have inferred masses
greater than 120 $M_{\odot}$. This corresponds to an absolute $K$ magnitude 
brighter than $-6.2$. For the stars
without $J$ magnitudes, they could have infrared excesses which push their
$Ks$ magnitudes above this. 

This is almost certainly the case for object F. Unpublished
spectra indicate that it has a spectrum marked only by lines of He~I at
2.06~$\mu$m, Br~$\gamma$, and anomalous features at 2.08~$\mu$m in 
emission and 2.10 in absorption 
(P. Conti and P. Crowther, private communication). The important point is
that no photospheric lines are seen. We can put an upper limit of 20
on its $J$ magnitude, for a minimum $J-H$ color of 4.2, and a maximum
Ks excess of $\sim 1.5$ magnitudes. 

One of the objects with inferred M $> 120$~$M_{\odot}$ has $J-H$ and 
$H-Ks$ colors that 
indicate it does not have a $Ks$ excess. It has an inferred 
absolute K magnitude
of $-7.14$, which is highly overluminous, even for a multiple of 3 objects. 
One possibility is that this object is slightly older than the surrounding 
stars. The stellar 
evolutionary models
for high mass stars predict that a 120 M$_{\odot}$ star will enter the 
supergiant phase at 1.7~Myr, and brighten by about 1.5 magnitudes at K.
An age spread of 1~Myr would explain this. Another less interesting possibility
is that the magnitudes are simply off due to difficulty in correctly 
characterizing the surrounding nebular emission.

\begin{table*}
\caption[]{Candidate Stars Associated with HII\ Regions}
\label{tab:uchii}
\begin{center}
\begin{tabular}{lllcccccc}\hline\hline
Radio Source & RA (2000) & Dec (2000) & $J$ & $H$ & $Ks$ & $J-H$ & $H-Ks$ & Inferred Mass (M$_{\odot}$)\\\hline
CC & 19:10:11.60 & 9:07:06.5 & $19.09\pm0.02$  & $15.78\pm0.02$ & $13.92\pm0.04$ & 3.31 & 1.86 & 56 \\
F  & 19:10:13.42 & 9:06:22.0 & ---    & $15.80\pm0.01$ & $11.94\pm0.02$ & ---  & 3.86 & $>120?$\\
J2 & 19:10:14.22 & 9:06:27.4 & ---    & $17.36\pm0.01$ & $15.45\pm0.02$ & ---  & 1.91 & 25 \\
O3 & 19:10:16.92 & 9:06:10.9  & $18.61\pm0.02$  & $15.54\pm0.02$ & $13.90\pm0.03$ & 3.07 & 1.64 & 46  \\
RQ & 19:10:10.76 & 9:05:18.2  & ---    & $19.18\pm0.03$ & $15.87\pm0.03$ & ---  & 3.31 & 78   \\
RQ & 19:10:10.93 & 9:05:15.9  & ---    & $17.33\pm0.02$ & $15.63\pm0.02$ & ---  & 1.70 & 18   \\
RQ & 19:10:10.96 & 9:05:17.7  & ---    & $18.57\pm0.03$ & $12.94\pm0.02$ & ---  & 5.64 & $>120?$ \\
RQ & 19:10:11.16 & 9:05:11.6  & $17.84\pm0.01$  & $15.51\pm0.02$ & $14.36\pm0.02$ & 2.34 & 1.14 & 21 \\
S  & 19:10:11.66 & 9:05:27.5  & $16.48\pm0.02$  & $14.16\pm0.02$ & $13.02\pm0.04$ & 2.32 & 1.14 & 46 \\
S  & 19:10:11.80 & 9:05:27.1 & ---     & $18.19\pm0.04$ & $15.23\pm0.05$ & ---  & 2.96 & 80 \\
S  & 19:10:11.83 & 9:05:28.3 & ---     & $17.21\pm0.03$ & $14.57\pm0.04$ & ---  & 2.64 & 84 \\
S  & 19:10:11.88 & 9:05:28.2 & ---     & $16.73\pm0.01$ & $14.40\pm0.02$ & ---  & 2.33 & 68 \\
South & 19:10:21.90 & 9:04:57.1 & --- & $17.60\pm0.02$ & $14.96\pm0.04$ & ---  & 2.64 &  68 \\
South & 19:10:22.08 & 9:05:00.0 & --- & $17.43\pm0.02$ & $14.54\pm0.03$ & ---  & 2.89 &  111 \\
South & 19:10:22.09 & 9:05:01.5 & --- & $19.02\pm0.04$ & $16.29\pm0.02$ & ---  & 2.73 &  36 \\
South & 19:10:21.97 & 9:05:04.2 & --- & $16.61\pm0.02$ & $14.54\pm0.03$ & ---  & 2.07 &  49 \\
South & 19:10:22.20 & 9:05:04.2 & --- & $18.23\pm0.03$ & $16.19\pm0.02$ & ---  & 2.05 &  19   \\
South & 19:10:21.90 & 9:04:58.1 & --- & $16.78\pm0.05$ & $14.74\pm0.04$ & ---  & 2.04 &  42 \\
South & 19:10:22.20 & 9:05:01.5 & --- & $18.06\pm0.03$ & $15.84\pm0.04$ & ---  & 2.22 &  28 \\
South & 19:10:21.80 & 9:05:03.3 & --- & $18.33\pm0.03$ & $14.34\pm0.05$ & ---  & 3.99 & $>120?$ \\
South & 19:10:21.88 & 9:05:04.3 & $20.07\pm0.04$ & $15.36\pm0.01$ & $12.74\pm0.03$ & 4.71 & 2.62 & $>120?$ \\
South & 19:10:21.75 & 9:05:04.3 & --- & $18.61\pm0.03$ & $14.88\pm0.05$ & ---  & 3.73 & $>120?$\\
South & 19:10:22.24 & 9:05:00.3 & --- & $18.46\pm0.01$ & $16.29\pm0.03$ & ---  & 2.17 & 20  \\
South & 19:10:22.02 & 9:05:00.3 & --- & $18.27\pm0.03$ & $16.11\pm0.05$ & ---  & 2.16 & 22  \\
South & 19:10:22.00 & 9:05:06.5 & --- & $18.56\pm0.03$ & $16.36\pm0.04$ & ---  & 2.20 & 20  \\\hline

\end{tabular}
\begin{list}{}{}
\item Notes: Names are from De~Pree~et~al.(1997).
Sources associated with RQ, S, and W49 South are likely 
to be multiple unresolved objects; nebular emission near these sources 
complicates the determination of the background and likely makes the errors in the photometry
larger than the quoted formal photometric errors. Masses are inferred from the 
extinction-corrected K magnitudes, a distance of 11.4~kpc, and $4\times10^{5}$~yr
stellar tracks from Lejeune \& Schaerer (2002) with solar metallicity and
enhanced mass loss. Objects with only $H$ \& $Ks$ magnitudes may have an unknown 
$Ks$ excess which could severely affect the inferred mass. Five objects 
have inferred masses above 120 M$_{\odot}$, which is likely due to a $Ks$ excess
and/or multiplicity. Errors of $10-20$\% in the mass estimate are expected simply
from the uncertainty in the distance.                   
\end{list}
\end{center}
\end{table*}

\section{Discussion}

\subsection{The W49A Star Cluster}

The virial mass of 
the W49 molecular cloud, $10^{6}$~M$_{\odot}$, puts it among the most 
massive in our galaxy (\cite{Setal01}). Our NIR 
observations cover the densest regions of this cloud, 
over a physical distance of 15~pc. There are ``fuzzy'' patches in 
our $Ks$ image from nebular emission, and these extend to the
Eastern, Western, and Southern edges of our field, indicating
that we have not fully sampled the star formation activity in 
the W49A cloud. There is also a peak in both the molecular gas density
(\cite{Setal01}) and the radio emission (\cite{BT01}) to the 
Northeast of our field.
   
What we {\it have} uncovered is a previously undetected 
massive stellar cluster (Cluster~1), and stellar sources 
associated with UC~\HII\ regions. Cluster~1 and the ``ring'' of 
UC~\HII\ regions are separated by only 2~pc in projection, meaning this differs
from a ``second generation'' as seen in 30~Doradus (\cite{Wetal99,Wetal02})
and NGC~3603 (\cite{Betal00,Netal02}). In the case of W49A,
when the OB stars powering the
UC~\HII\ regions emerge, the region encompassing both Cluster~1 and
the Welch ring of UC~HII regions 
will appear to be the ``core'' of the star cluster.
The projected geometry of the 
region is highly suggestive of triggering; the ``ring'' of UC~\HII\ regions
is at the border of the ionized bubble surrounding Cluster~1.

What does the core of Cluster~1 hold? Given the high internal extinction, 
we are likely to be incomplete in our near-infrared 
census of star formation and therefore a
total mass or density estimate. 
The core is crowded; high spatial resolution observations are needed to
accurately determine the stellar density in the core.
 Taken at 
face value and without correcting for the large extinction,
the cluster core appears to be 
significantly less dense 
than the Arches, NGC~3603, or 30~Doradus. If it is truly less dense,
then the different formation environment of  
W49A, at a Galactocentric distance of 8~kpc, may be an important 
clue for understanding the processes
which drive clustered star formation.

The subclustering phenomenon is useful to describe the 
star formation pattern in the W49A molecular cloud. 
When the cloud has ceased forming stars, the resulting 
stellar group will likely be called a 'cluster'. At the
time of current observation, the massive star formation 
does not appear to be distributed uniformly throughout the region,
or with a radial dependence relative to a cluster 'center'. 
Rather it is better described as occuring in 'subclusters'. 
In this sense we could count $4-5$ subclusters within $\sim 13$~pc
using the combined NIR and radio
observations: Cluster~1, the (Welch) ``ring'' of UC~\HII\
regions, W49A South, the RQ complex, and perhaps the CC source.
We speculate that star formation within a subcluster is essentially 
synchronized, and a massive star cluster is a collective of
several (or many) subclusters.  

\subsection{Total Stellar Mass}

We can make an estimate for the total stellar mass of the W49A star
cluster by counting stars with masses greater than $20$~M$_{\odot}$ 
and using a Salpeter slope for the mass function. We take upper and
lower mass limits as 120~M$_{\odot}$ and 1~M$_{\odot}$, respectively. 
For Cluster~1, we
find 54 stars within $45 \arcsec$, implying a total mass of 
$\sim 1 \times 10^{4}$~M$_{\odot}$. In our entire field, we 
count 269 stars with masses $\ge 20$~M$_{\odot}$, implying a
total mass of $5-7 \times 10^{4}$~M$_{\odot}$. The stars we
have identified as massive stars are
certainly contaminated by background objects,
but we are also certainly incomplete in our census due to extinction and
angular resolution. The extent to which these effects cancel 
each other out (or not) is unknown. Even if the stellar 
mass estimate for W49A is a factor of 2 too high, W49A is as or 
more massive than 
any known young Galactic star cluster. We also note that it is possible,
perhaps even likely, that we have not 
yet detected the most massive young star clusters in the 
Milky Way (e.g. \cite{H03}). 

It is important to note that this is a {\it lower limit} 
to the final stellar mass,
as there is circumstantial and direct evidence for ongoing star 
formation in this region. There is abundant molecular gas,
and hot cores near the ring of UC~\HII\ regions (\cite{Wetal01}, \cite{MGD04}). 
The densest region of the molecular cloud is north of the ``ring'' of
UC~\HII\ regions, which is completely extincted even in our $Ks$ image.
This is the most likely place for massive stars in earlier stages of
formation than currently probed with existing observations.
What we observe in W49A is a region with massive stars at various
evolutionary stage, from hot cores to UC~\HII\ regions to naked 
OB stars, similar to W43 (\cite{Betal99}; \cite{Metal03}) and
the significantly less massive W75N (\cite{Setal03}).

%\subsubsection{Comparison to Embedded Extragalactic SSCs}

%In recent years there has been a surge of interest in embedded
%extragalactic super star clusters (SSCs) (e.g. \cite{Jetal01,PS02,JIP03,TB04}
%as the youngest known stages of massive star cluster formation.
%As W49A is the most massive embedded cluster in the known Milky Way, it
%has been used as a comparison object to extragalactic SSCs. In
%most such comparisons, W49A is $1-2$ orders of magnitude {\it less} luminous
%than these ``ultra-dense H~II regions'', or UD~HII regions. The results
%presented here suggest that the ionizing photon luminosity of W49A 
%is underestimated by at least a factor of 2, and possibly a factor of 4. 
%Unless the fraction of photons absorbed by dust varies widely
%from region to region, this suggests that the ionizing luminosities
%and thus the stellar masses 
%of UD~HII regions are larger than inferred from radio observations
%alone. 

\section{Conclusions}

Here we have presented a more comprehensive investigation into our previous 
discovery of stellar clusters in the giant radio HII\ region W49A 
(\cite{AH03}). 
Our observations clearly show a massive star cluster 
adjacent to the UC~\HII\ regions (2~pc distant).
This means that the W49A region began forming stars 
earlier than previously thought, and that the
ultra-compact HII\ regions which have been long-known to radio astronomers
are not the first generation of massive stars. 
We use these data to estimate a total stellar mass in this region of 
$5-7\times10^{4}$~M$_{\odot}$, and a total mass for Cluster~1 of
$1\times10^{4}$~M$_{\odot}$. Since molecular gas is
abundant, this is a lower limit to the final stellar mass of the cluster.

With these observations, W49A joins the list of Galactic giant
radio \HII\ regions with the coexistence of two or more phases of 
massive star formation. This means that the formation of 
massive stars is not completely synchronized, but that there 
is some spread in age. 
The magnitude of the spread could be investigated with spectra of 
the relatively unembedded massive stars and the 
lifetime of UC~\HII\ regions, although this lifetime is only poorly known.
With the current observations and the
presence of dense molecular gas in the central few pc, a reasonable
guess for the age spread is 1~Myr.

The last point we would like to make is that the subclustering 
phenomenon is essential for the description of star formation in 
the W49A molecular cloud, at least as traced by the massive stars. 
However, there also exists evidence for
subclustering in lower-mass star-forming regions (\cite{LAL96,Tetal00}),
which is reproduced in star formation simulations (\cite{BBV03}).
Possible examples of subclustering in
extragalactic star clusters are: SSC-A in NGC~1569 and NGC 604 in M33. 
SSC-A in NGC~1569 has a
stellar concentration with red super giants and another with 
Wolf-Rayet stars (\cite{Getal97,DMetal97,Hetal00,Oetal01}).
The massive stars in NGC 604 are subclustered, but the
region itself it is of sufficiently 
low density to be termed a Scaled OB Association (SOBA) rather
than a star cluster (\cite{M-A01}). The applicability of the subclustering 
description to other young massive Galactic star clusters remains to be seen,
but we conclude that it is a useful concept to describe and understand massive
star formation in the W49 GMC.

\acknowledgements{NH acknowledges and thanks the European Southern 
Observatory (ESO) Studentship Programme which provided support during
the early stages of this work.
We are pleased to acknowledge Miguel Moreira for discussions and
assistance with the observations, Robert Simon for providing molecular
line data on W49's giant molecular cloud, where the clusters are
embedded, and Chris De Pree for providing radio continuum data of the
\HII\ regions associated with W49A.
}

% \begin{figure}
% \plottwo{f2a.eps}{f2b.eps}
% \caption{This is an example of a multipart figure with a long figure caption 
% that must be set as a paragraph.  The processor has to buffer the text of the
% caption, so it is good not to be too wordy, but that would make for
% poor communication as well.\label{fig2}}
% \end{figure}

\end{document}